# An Anomaly in the Angular Distribution of Quasar Spectra

Michael J. Longo[1]
Department of Physics, University of Michigan, Ann Arbor, MI 48109, USA

**Abstract**
Quasars provide our farthest-reaching view of the Universe. The Sloan Survey now contains over 100,000 quasar candidates. A careful look at the angular distribution of quasar spectra shows a surprising "bullseye" pattern on the sky toward $(\alpha, \delta) \sim (190°, 0°)$ for all wavelengths from UV through infrared. The angular distribution of the shift in the UV suggests a large peculiar velocity $v_p$ toward that direction. However, the size of the shift would indicate a $v_p \sim 0.2\ c$, which is two orders of magnitude larger than measures of our peculiar velocity from nearby galaxies and cosmic microwave background (CMB) measurements. The angular pattern and size of the shift is very similar for all wavelengths, which is inconsistent with a Doppler shift. The shift is also too large to explain as a systematic error in the quasar magnitudes. The anomaly appears to be a very large hotspot in the Universe. Its direction is close to that of the reported anomalies in the CMB, the so-called "axis of evil". The angular pattern of the shift and its redshift dependence are consistent with the existence of an expanding bubble universe in that direction, which could also explain the CMB anomalies.

## 1. Introduction

Quasars, by virtue of their enormous luminosities, offer our farthest-reaching picture of the Universe. By now, the Sloan Digital Sky Survey (SDSS) [1] has found over 100,000 quasar candidates [2] out to redshifts $z$ beyond $z = 5$. These have well measured redshifts based on spectral lines and relative magnitudes for 5 contiguous wavelength bands ranging from 400 nm to 1000 nm [3]. These correspond to ultraviolet $U$, green $G$, red $R$, infrared $I$, and far-infrared $Z$. The filter magnitudes can be thought of as a measure of the apparent brightness of the quasar in each color band. Quasar candidates are selected via their nonstellar colors in *UGRIZ* broadband photometry [4]. The selection is based upon measurements that have been corrected for Galactic extinction using the maps of Finkbeiner, Davis, and Schlegel [5].

The distribution of the quasar sample in redshift is highly biased because the selection criteria are $z$ dependent. There is also the natural falloff at large $z$ as the quasars become fainter.

---
[1] email: mlongo@umich.edu

However, it is important to note that the selection criteria do not depend explicitly on the angular coordinates: right ascension $\alpha$ and declination $\delta$. The filter magnitudes have been corrected for atmospheric effects, and the magnitudes used in this study were corrected for Galactic extinction [5]. Thus, a dependence on $(\alpha, \delta)$ of the continuum spectra from the filters is not expected. The technique used here is therefore to find the $z$ dependence of the entire sample and subtract the magnitude averaged over the entire sample at a given $z$ from that for each quasar to look for a possible systematic dependence on $(\alpha, \delta, z)$. As we shall discuss below, there is a significant shift in all the filter magnitudes, ~0.25 magnitude brighter, for $\alpha \sim 190°$ and $\delta \sim 0°$. It is also important to note that because the quasar redshifts were determined from spectral lines, a large peculiar velocity $v_p$ toward this direction would only cause a shift in the $z$ assigned to each quasar that is $\sim v_p/c$. The filter magnitudes would still show the same angular dependence characteristic of a rapidly moving observer.

There are several ways of determining the peculiar velocity of our Galaxy relative to the surrounding universe. Techniques using the bulk flow of galaxies are limited to $z \leq 0.05$. For example, Nusser and Davis [6] find a peculiar velocity about 257 km/s, or $v_p/c \sim 8.6 \times 10^{-4}$, toward $(\alpha, \delta) = (158°, -46°)$ within $z = 0.033$. Dai *et al*. [7], using Type 1A supernovae, find a bulk flow with $v_p \sim 188$ km/s along $(\alpha, \delta) = (158°, -46°)$ for $z < 0.05$, while the data for $z > 0.05$ show no evidence for bulk flow. On the other hand, Kashlinsky *et al*. [8], using the bulk flow of X-ray luminous clusters of galaxies, find $v_p \sim 1000$ km/s along $(\alpha, \delta) \sim (175°, -33°)$ out to $z \sim 0.25$. The CMB dipole found by WMAP [9] indicates that the Solar System is moving at 368 km/s, or $v/c = 1.23 \times 10^{-3}$, relative to the observable Universe in the direction $(\alpha, \delta) = (167.9°, -6.9°)$ with an uncertainty <0.1°.

It is difficult to reconcile these different measures of our peculiar velocity. The discrepancy between the CMB dipole direction and the bulk flow measurements could be due to a relatively local anisotropy in the mass distribution. For example, Tully *et al*. [10] suggest that a large component of the bulk flow of our Galaxy is due to our movement away from an extremely large Local Void. In any case, none of these measures of our peculiar velocity give a $v_p/c \gtrsim 3 \times 10^{-3}$.

## 2. Analysis

For this analysis, all 105,783 of the quasar candidates from the SDSS Quasar Catalog V, DR7 [2] were used. Since the density of quasar candidates falls off rapidly beyond $z \sim 2.2$, it was restricted to $z < 2.2$, which corresponds to 6600 $h^{-1}$ Mpc. Redshifts less than 0.2 were also excluded because the spectra were varying rapidly with $z$ due to the selection criteria. This left approximately 82,500 quasars. Of these, 75,300 were in the right ascension range $110° < \alpha < 255°$ and declinations -10°< $\delta$ <70°. The remaining 7,200 were in -50°< $\alpha$ <65° in 3 narrow declination bands near -8°, 0°, and 15°.

Files with the ($\alpha$, $\delta$) coordinates, redshift, redshift errors, and *UGRIZ* magnitudes corrected for Galactic dust extinction were downloaded from the SDSS website. Much of the subsequent discussion refers to the *U*ltraviolet magnitudes, though the other filter magnitudes showed a similar behavior. Figure 1 shows the smoothed *U* and *Z* filter magnitudes vs. redshift for all the quasars in the sample. An estimated absolute *U* magnitude is also shown.

For each quasar the deviation of the *U* magnitude from the average *U* magnitude for that redshift was calculated as

$$\Delta U = U(\alpha,\delta,z) - \bar{U}_{int}(z) \qquad (1)$$

where $\bar{U}_{int}(z)$ is the average *U* for that redshift, as interpolated from the curve in Fig.1. As defined, a negative $\Delta U$ corresponds to the quasar being brighter than average in that band.

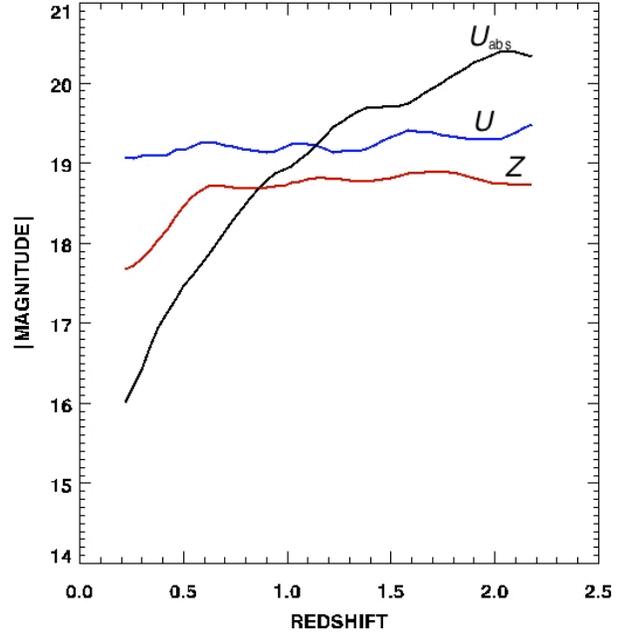

Figure 1 – Smoothed *U* and *Z* magnitudes for the entire sample vs. redshift. The black curve shows the approximate absolute *U* magnitude. Its sign has been dropped for convenience.

The sample was then divided into slices in declination. Polar plots of declination vs. *z* for two of the $\delta$ slices are shown in Fig. 2. The points give the ($\alpha$, *z*) position of each quasar. Their colors indicate $\Delta U$ for that quasar. The colors range from violet through blue, green, orange, and red, with violet representing the largest blueshift (most negative $\Delta U$), red the largest redshift, and green for *U* approximately equal to the average. Color contours are also shown in Fig. 2. These were generated by binning the $\Delta U$ for each $\delta$ slice into 16 bins in $\alpha$ and 8 bins in *z* for a total of 128 bins with typically ~100 entries per bin. This array was then smoothed and contours plotted. The contours range from $\Delta U$ = -0.2 (violet), -0.1 (blue), 0.0 (black), 0.1 (orange), and 0.2 (red). The contours and points in the hemisphere toward $\alpha$ = 0° correspond to the entire $\delta$ range there, because of the limited range of declinations in the sample in that hemisphere. Thus they are the same in both $\delta$ slices.

The systematic blueshift toward $\alpha$ =180° is especially apparent for the 10°<$\delta$<20° slice. There is no sign of it in the other hemisphere. Six $\delta$ slices in the left hemisphere were studied in 10° increments in $\delta$ from 0° to 60°. These are statistically independent and all show the same systematic blueshift which appears to be greatest in the 0°<δ<10° and 10°<δ<20° slices, gradually decreasing toward δ = 60°.

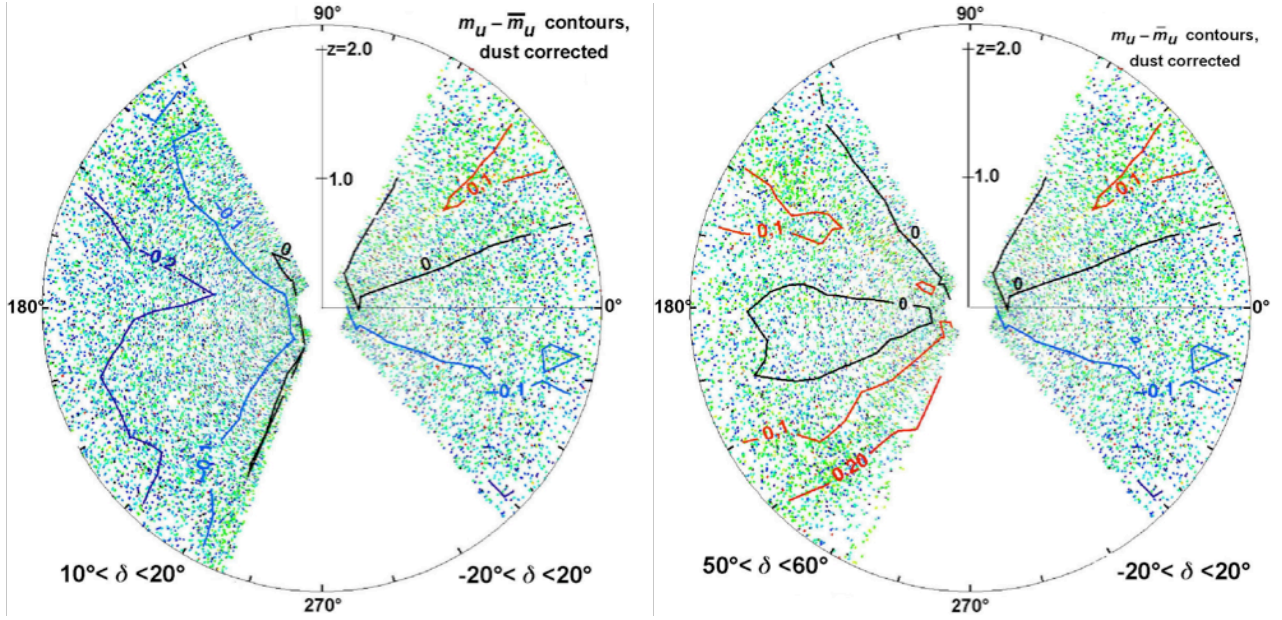

Figure 2 – Polar plots of quasar UV band color shift vs. redshift and right ascension for two declination slices. The overall $z$ dependence has been removed. The contours, described in the text, range from $\Delta U$ = -0.2 (violet), -0.1 (blue), 0.0 (black), 0.1 (orange), and 0.2 (red). The dots indicating the quasar positions have a similar color correspondence, ranging from violet-blue-green-orange-red. The gaps toward 90° and 270° are due to the obscuration by the Galactic disk. The hemisphere toward 0° is the same in both plots since the SDSS coverage there is limited to 3 small $\delta$ ranges between -15° and 15°. For the hemisphere toward 180° the $\delta$ ranges are 10° to 20° for the left plot and 50° to 60° for the right plot. The blueshift of the spectra toward 190° for small $\delta$ is apparent in the left plot.

A more quantitative appreciation of the blueshift toward $\alpha$ = 190° can be gotten by looking at the variation of $\langle \Delta U \rangle$, the average $\Delta U$ for each bin in the 128 $\alpha$, $z$ bins described previously. This is shown in Fig. 3 for the slices 0°<δ<10° and 50°<δ<60°. The 8 sets of curves show the variation of $\langle \Delta U \rangle$ with $\alpha$ for each of the 8 $z$ bins. The red (dash-dot) curves are for the right hemisphere in Fig. 2 translated by 180° to allow an easy comparison, and are the same for both δ slices. The error bars are calculated from the standard deviation of the $\Delta U$ for the <u>entire</u> quasar sample divided by the square root of the number of entries in the bin. For 0°<δ<10°, all of the $z$ ranges above $z$=0.7 show a dip of about 0.25 in $\langle \Delta U \rangle$ for 150°≲ $\alpha$ ≲ 220°. This dip is much less apparent for the 50°<δ< 60° plots, and is not present in the dash-dot curves for the other hemisphere. The angular correlations, as well as the dips in Fig. 3, disappear if the $\alpha$'s and δ's are scrambled among the quasars. Note that all of these curves are statistically independent. The other 4 δ slices, which are also statistically independent, appear to show a similar behavior with a larger dip at small declinations. Thus the statistical significance of the effect is overwhelming.

The angular dependence of the effect can be seen by looking at the ($\alpha$, δ) dependence within a spherical shell containing a small range of redshifts. A plot of $\langle \Delta U \rangle$ vs. declination and right ascension for a spherical shell with 1.7 < $z$ < 1.95 is shown in Fig. 4. The overall $z$ dependence has been removed. As before the contours are based on the smoothed $\langle \Delta U \rangle$, in this case for 16 $\alpha$ and 10 δ bins. The contours range from $\langle \Delta U \rangle$ = -0.001 (violet), -0.0005 (blue), 0.0 (black), 0.0005 (orange), and 0.2 (red). Because of the small range in $z$, 0.25, these $\langle \Delta U \rangle$ have a much smaller range than they do for the ($\alpha$, $z$) plots of Fig. 2. The "hotspot" (blue-violet) toward (200°, 10°) is apparent. It covers an angular range ~90° in $\alpha$, and at least 30° in δ. The $\Delta U$'s for the entire sample within this shell must sum to 0, so that the hotspot appears to be surrounded by a colder (redder) region. In other words, we are only looking at $\langle \Delta U \rangle$ <u>differences</u> over the range of ($\alpha$, δ) covered by SDSS. Because the hotspot includes a large fraction of the total sample, its contrast with the surroundings is reduced somewhat by the averaging procedure.

Plots of the ($\alpha$, δ) dependence for the 7 other, statistically independent, redshift shells studied all show a very similar behavior. The angular range and magnitude range of the hotspot is nearly the same for all $z$ ≳ 0.5; for smaller $z$ it appears to diminish in size and magnitude range. Again, the angular correlations disappear if the $\alpha$'s and δ's are scrambled among the quasars. The declination range of the hotspot may be truncated by the lack of coverage of SDSS below δ ~ -5°, and it probably extends farther toward negative declinations. There are also hints of large-scale structure in the quasar distribution, particularly in the region just above the $\langle \Delta U \rangle$ = 0 contour. If real, its scale is far larger than any previously seen.

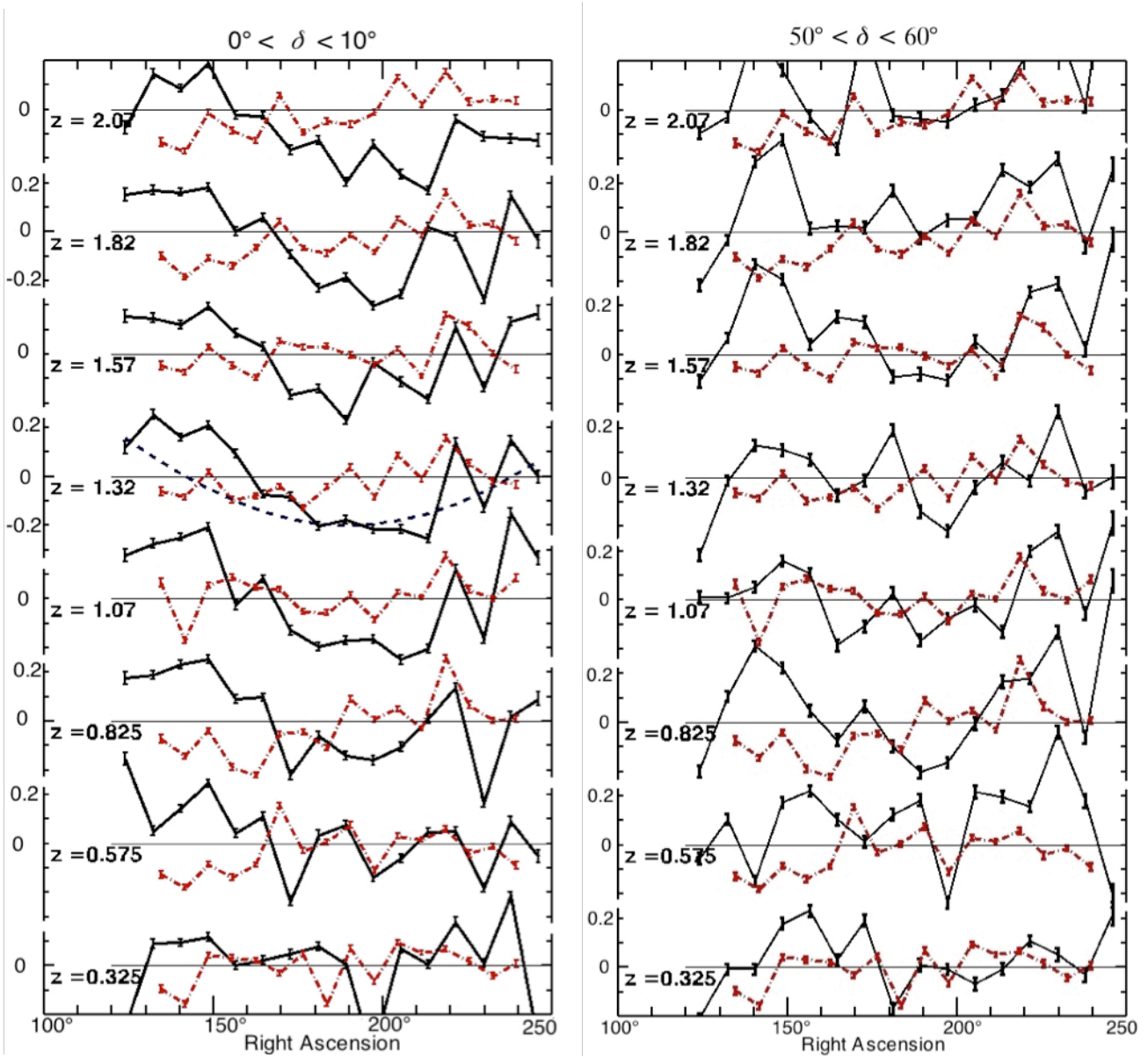

Figure 3 – The variation of $\langle \Delta U \rangle$ with $\alpha$ for each of the $z$ bins for two of the $\delta$ slices. Negative $\langle \Delta U \rangle$ correspond to a higher brightness in the blue. The red (dash-dot) curves are for the right hemisphere in Fig. 2 translated by 180° to allow an easy comparison, and is the same in both plots. The dashed curve at $z = 1.32$ in the left plot shows the expected cosine variation for motion toward $\alpha = 190°$.

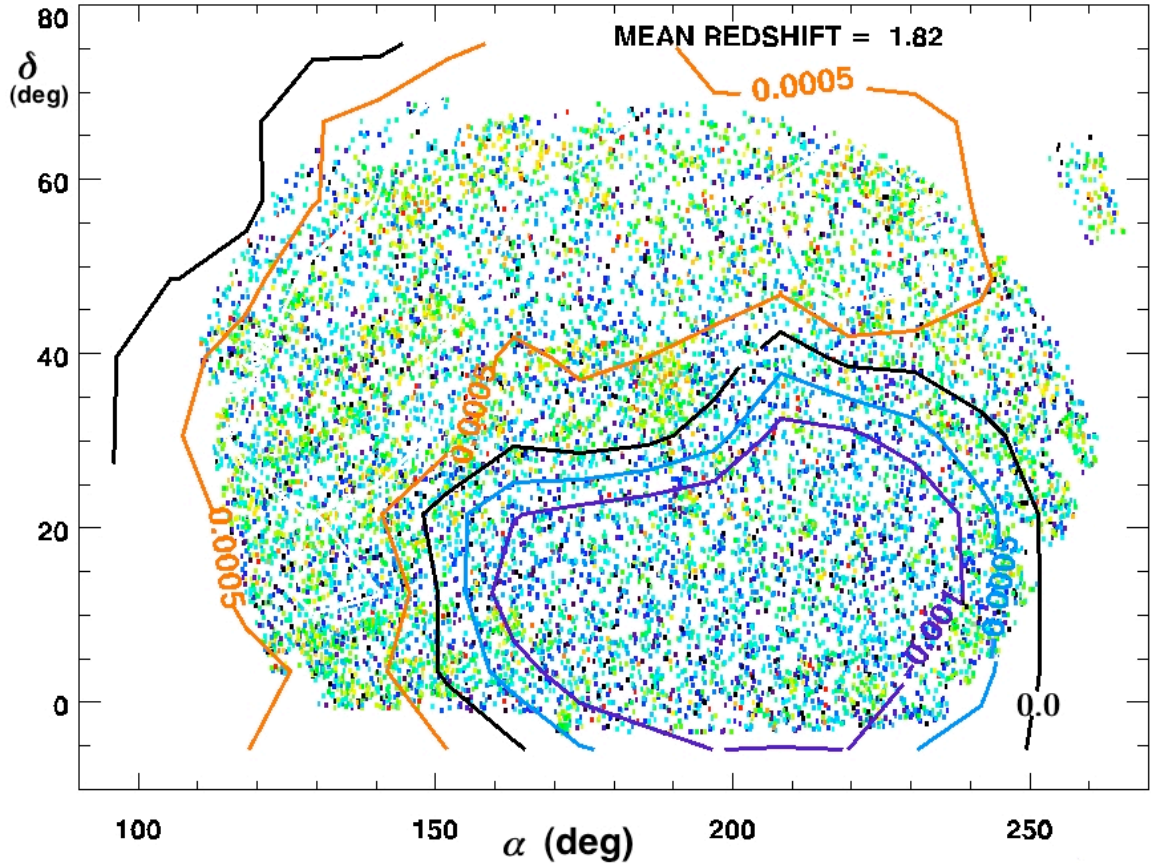

Figure 4 – Plot of $\langle \Delta U \rangle$ vs. declination and right ascension for a spherical shell at a radius $1.7 < z < 1.95$. The overall $z$ dependence has been removed. The contours range from -0.001 (violet), -0.0005 (blue), 0.0 (black), 0.0005 (orange). The dots indicating the quasar positions have a similar color correspondence, ranging from violet-blue-green-orange-red. The "hotspot" (blue-violet) toward (200°, 10°) is apparent. The $\Delta U$'s for the entire sample within this shell must sum to 0, so that the hotspot appears to be surrounded by a colder (redder) region. The same "bullseye" pattern appears for all redshift ranges above $z$~0.5 and for all 5 filter bands.

## 3. Discussion

The shift in the UV continuum spectra of the quasars toward $(\alpha, \delta) \sim (190°, 0°)$ is ~0.25 in magnitude. The quasar redshifts are determined from spectral lines so their $z$ would be systematically shifted, making them appear to be slightly closer toward $(\alpha, \delta) \sim (190°, 0°)$. In the Doppler shift scenario, this shift would indicate a very high velocity toward that direction. The shift in magnitude ~0.25 can be related to the shift in $z$ using the $U_{abs}$ dependence shown in Fig. 1. For that curve, the typical slope $\Delta U_{abs}/\Delta z$ for redshifts between 1 and 2 is about 1.5 magnitudes per redshift. Thus a shift in magnitude $\Delta U$ of 0.25 corresponds to a $\Delta z \sim 0.17$. The non-relativistic Doppler shift for a velocity $v_p$ is

$$\Delta z \equiv \frac{\lambda_O - \lambda_S}{\lambda_S} = \frac{v_p}{c} \cos\theta \qquad (2)$$

where $\lambda_O$ and $\lambda_S$ are the wavelengths seen by the observer and source respectively and $\theta$ is the angle between the quasar and our velocity. The apparent blueshift therefore implies a velocity ~$0.17\,c = 5 \times 10^4$ km/s toward $(\alpha, \delta) \sim (190°, 0°)$. This is in approximately the same direction as the CMB dipole toward $(167.9°, -6.9°)$, but it is far greater than the 368 km/s found for the CMB [9]. (However, the CMB redshift corresponds to a $z\sim 1000$ at decoupling, so these two measures of peculiar velocity can be somewhat different.) It is also far larger than the bulk flow measurements [6, 7, 8] which are typically toward $(\alpha, \delta) \sim (158°, -46°)$ and correspond to $z \sim 0.1$. The direction of the blueshift is also close to that of the so-called "Axis of Evil", a name coined by K. Land and J. Magueijo [11] to describe the anomalies in the low multipoles of the CMB at $(\alpha, \delta) \sim (173°, 4°)$.

However, the overall pattern of this apparent blueshift is **not** consistent with a Doppler shift due to our peculiar velocity toward $(\alpha, \delta) \sim (190°, 0°)$. The above discussion was for the $U$ (ultraviolet) filter band. The $G, R, I,$ and $Z$ band spectra were found to have very similar shifts in filter magnitudes, both in size and for their $(\alpha, \delta, z)$ dependence. There is also no evidence for a redshift in the opposite hemisphere that would be characteristic of a large Doppler shift away from it. The angular pattern of the magnitude shifts cannot be explained as a systematic error in the filter magnitudes. All of the magnitude shifts along the $z$ direction are ~0.25, while the quoted uncertainties in the magnitudes are ~0.03 for each quasar [2]. It is hard to conceive of an angle-dependent systematic error that is similar for all 5 filters that would mimic the observed angular pattern over the entire $z$ range and also almost coincide with the CMB anomaly axis.

Since the Doppler shift interpretation of the shifts is not tenable, we must consider other explanations. A Doppler shift due to our movement toward that region would shift the radiation blue-ward, which would cause the blue component to increase at the expense of the infrared, while the observed shift occurs over the whole spectrum. This rise in the overall spectrum of the continuum radiation must be due to a higher temperature of the quasars in that region or of their environment. The hotspot is very large with an angular size ~50°. The hotspot's features could be explained if another universe, located around $(\alpha, \delta)$ ~ (190°, 0°) and $z$ ~ 2, is expanding into our Universe. This bubble universe could have originated somewhat after ours but within it. We would be seeing it at a lookback time ~10 Gyr when it was smaller and hotter; hence that region appears brighter and hotter. That region would also be denser and the overlapping region would be heated by the shock wave generated as the bubble expands into our universe, driving gas and magnetic fields along with it. The higher gradient of the effect along the $z$ direction compared to that along the $(\alpha, \delta)$ direction (i.e., the difference in range of the contour levels in Fig. 2 vs. those in Fig. 4) could be due to the fact that as we look farther in $z$ we are seeing the bubble at still earlier times when it was much hotter. There would also be a Doppler blueshift of the quasars in the expanding bubble due to their motion toward us, which would slightly distort the $z'$s that we observe. This bubble scenario appears to be the only one that is consistent with the overall $(\alpha, \delta, z)$ and wavelength dependence of the shift. It would also be consistent with the apparent lack of any shift in the hemisphere toward $\alpha = 0°$. It could explain the CMB anomalies, and the higher density of the overlapping region could help explain the bulk flow, which appears to be directed roughly in the same direction.

There has, of course, been considerable discussion of multiverses and bubble universes within or outside of our own universe. S. Feeney *et al*. [12] discuss possible observational tests to search for evidence of bubble universes. The "bullseye" CMB temperature modulation they show in their Fig. 1, top left, looks remarkably like that in Fig. 4 for the quasar magnitude shifts except for the incidental color inversion . If verified, this result would be the first strong evidence for another universe. It also shows that a careful look at quasar spectra can be used as a thermometer to look for hotspots and temperature anomalies in the Universe out to $z > 2$.

The confirmation of an expanding bubble universe toward $(\alpha, \delta)$ ~ (190°, 0°) will have profound implications for all of cosmology. Strict homogeneity and isotropy would no longer be the norm on the largest scales. At first glance, the bubble universe seems much like ours with a similar mass distribution and size compared to ours some 10 Gyr ago. It will be fascinating to explore it further.